\documentstyle[12pt,cite,epsf]{article}
\title{Contribution of Kaluza-Klein modes to vacuum energy in 
models with large extra dimensions $\&$ the Cosmological constant}

\author{Abhinav Gupta\thanks{E--mail : abh@ducos.ernet.in, 
abhinav@physics.du.ac.in} \\
	{\em St. Stephen's College,} \\
	 {\em University of Delhi, Delhi-110 007, India.}}

\begin{document}
\maketitle
\begin{abstract}
 In this paper, the generation of topological energy in models
 with large extra dimensions is investigated. The origin of this energy
 is attributed to a topological deformation of the standard Minkowski vacuum
 due to compactification of extra dimensions. This deformation
 is seen to give rise to an effective, finite energy density due to
 massive Kaluza-Klein modes of gravitation. It's renormalized value  is seen
 to depend on the size of the extra dimensions instead of the UV cut-off of
 the theory. It is shown that if this energy density is to contribute to
 the observed cosmological constant, there will be extremely stringent
 bounds on the number of extra dimensions and their size.
\end{abstract} 
\vskip 1cm
\begin{section}*{Introduction}
\indent The Einstein equations with the cosmological constant are given by
\begin{equation}
R_{\mu\nu}-{1\over{2}}g_{\mu\nu}R=g_{\mu\nu}\Lambda-8\pi GT_{\mu\nu}
\end{equation}
where $\Lambda$ is the cosmological constant, or vacuum energy. Modulo
 the factor of $8\pi G$, this constant is identical to the stress tensor 
 associated with vacuum. It can be seen on grounds of covariance that
  the vacuum expectation value of the stress tensor is of the form \cite{zeldo}
\begin{equation}
\langle T_{\mu\nu}\rangle=\rho_{vac}\,g_{\mu\nu}
\end{equation}
\indent No known symmetry forbids a cosmological term in Einstain equations.
Based on observations, it is seen that in suitable units, $\Lambda$ is at
 most of the order of $10^{-47}~$ (GeV)$^{4}$. At the same time, the vacuum energy
 evaluated with an ultraviolet cut-off at the Planck scale $M_{pl}\sim 
10^{19} $ GeV gives
\begin{equation}
\langle E_{vac}\rangle \sim {1\over{2}}\int^{M_{pl}}_{0} {d^3k\over{(2\pi)^3
}}\sqrt{k^2+m^2}\sim M^{4}_{pl}
\end{equation}
which is about 120 orders of magnitude higher.
  Attempts have been made to
 account for this discrepancy \cite{wein,varun} and it has been realized
 that even if 
 the cut-off is taken at the Supersymmetry (SUSY) scale, 
 the discrepancy is still about 60 orders of magnitude. It
is seen that an  ultraviolet cut-off
 of the order of $10^{-3}$ eV is required to account for the observed value 
 of the vacuum energy. Therefore, even very low energy quantum effects at
 sizes much larger than the size of atoms lead to untolerable contributions
 to the cosmological constant.\\
\indent Several suggestions such as SUSY, scalar dilatons, 3-form fields
 and quintessence have been made, but no satisfactory resolution is
 seen to emerge (for a review, see \cite{ellwanger}).\\
\indent Recently, there have been developments in Kaluza-Klein theories that
 claim to remove the so called hierarchy problem between the Electroweak
 and Planck scales \cite{nima,randall}. In these recent models, it is
 assumed that all matter is confined to a four dimensional hypersurface,
 a 3-brane, in a higher dimensional spacetime. Gravity, unlike
 matter,  is assumed to propagate in all dimensions. There are essentially
 two models and their variants. The first model, the ADD model\cite{nima}
 , assumes
 the existence of n large extra dimensions with size 
 of the order of a millimeter and a bulk scale of gravitation of the order 
 of a TeV, instead of $M_{pl}$. The metric of spacetime is a direct product
 of the usual 4-dimensional spacetime and the compact extra dimensional space.
Due to the large volume of the compact space,
 gravitation on the 3-brane is diluted down to the observed Planck scale, 
 instead of a TeV scale, which is the gravitational scale of the bulk
 spacetime. The large size of the extra dimensions removes the 
 hierarchy between the two scales and is not in conflict 
 with micro-experiments, since matter cannot directly probe these extra
 dimensions, being stuck on the 3-brane. The most studied model from a 
 cosmological point of view is the RS model \cite{randall}
and it's variants. In this model,
 there is only one extra dimension and one or more branes. This model
 differs from the ADD model in that spacetime is not a direct product of
 the four dimensional spacetime and the extra dimensional manifold (as is
 the case in the ADD scenario), but has a warp factor multiplying the
 four dimensional metric. This warp factor is a function of the extra 
 dimension. There are cosmological constant terms on the 3-branes and in 
addition, a bulk cosmological constant. The 5-dimensional equations
 are solved to obtain the 4-dimensional effective gravitational 
constant and the effective induced 4-dimensional cosmological constant.
However, fine-tuning is required to get the effective cosmological
 constant to agree with the observational value. For a recent review on various
 approaches used to solve the cosmological constant problem, see 
\cite{ellwanger}\\
\indent In the ADD model, the vacuum solution is taken to be the 
 standard bulk Minkowski
 metric, with the n extra dimensions compactified on an n-torus. A torus
 is obviously a zero curvature manifold. However, it is topologically
 very dissimilar to a situation in which the extra dimensions are non-compact.
  As a result, the vacuum of the former manifold is dissimilar
 to the vacuum of the latter, non-compact manifold. It is a standard
 result in quantum field theory that if a quantum field propagates in all
 the dimensions, the vacuum energy associated with it will be different 
 in the two topologically dissimilar manifolds considered above.
 To get a finite, renormalized value for the vacuum energy associated
 with the compact manifold, the non-compact contribution is subtracted
 out. (For a 
detailed discussion, see \cite{birrel}). This is 
equivalent to stating that the absolute zero of energy
 is taken to be a state in which all dimensions are non-compact, and the
 metric Minkowskian. 
 In the ADD model, the
 gravitational field is the only field which propagates in all the dimensions,
 with the matter fields being forever stuck on the 3-brane. 
 Even in the so-called vacuum state, since the extra dimensions are compact,
 there will be a vacuum stress due to compactification, giving rise to
 zero point vacuum fluctuations which after renormalization will be shown
 to be finite. It will be seen that this net vacuum energy is nothing
 but the renormalized zero-point energy of all the massive Kaluza-Klein modes
 associated with the gravitational field. In the following, this vacuum
 energy is explicitly
 calculated for two extra compact dimensions, and the result generalized 
 to $n$ extra dimensions. It is seen that if this vacuum energy is to
 contribute to the observed cosmalogical constant, this imposes serious
 constraints on the number of extra dimensions and their size.\\ 
\end{section}
\begin{section}*{A Toy Model}
We begin by studying a toy model in 1+1 dimensions, in which the single
 spatial dimension is compact (See \cite{birrel}) and  spacetime is $R^{1}\times S^{1}$. The spatial
 points $x$ and $x+R$ are identified, where $R$ is the periodicity length
 or 'circumference' of the spatial section. Consider a massless scalar
 field propagating in this spacetime. Then, the effect of the compactification
 is to restrict the modes of propagation of the scalar field to the form
\begin{equation}
\phi_{k}={1\over{\sqrt{2R\omega}}} e^{\imath(kx-\omega t)}
\end{equation}
 where
 $k=(2m\pi)/R$ and $\omega=|k|$, $m$ being an
 integer. The field $\phi$ is expanded in
 terms of creation and annihilation operators and the modes $\phi_k$
 as
\begin{equation}
\phi(t,x)=\sum_{m}[a_m\phi_{m}+a^{\dagger}_{m}\phi^{*}_{m}]
\end{equation}
\\
\indent The vacuum expectation value of the energy density is calculated
 to be
\begin{equation}
\langle 0_R |T_{00}|0_R\rangle={2\pi\over{R^2}}\sum_{m=0}^{\infty}m \label{eq1}
\end{equation}
 where $|0_R\rangle$ is the vacuum state for this compact manifold and
 is defined by $a_m|0_R\rangle=0$.
 As expected, this is infinite. However, we renormalize it by demanding 
 that the vacuum energy associated with the manifold $R^1\times R^1$
 should be zero, which is the standard Minkowski manifold in the limit
 $R\longrightarrow\infty$. This renormalization implies that we subtract
 the vacuum contribution as $R\longrightarrow\infty$ from the 
 expression given in eqn.(\ref{eq1}),i.e., the renormalized energy density of
 the compactified vacuum is given by
\begin{equation}
\langle 0_R|T_{00}|0_R\rangle_{ren}=\langle 0_R|T_{00}|0_R\rangle
-\langle 0_R|T_{00}|0_R\rangle\vert_{R\longrightarrow\infty}
\end{equation}\\
\indent This can be achieved by standard regularization techniques 
(see\cite{birrel})
 in which the vacuum expectation value is regularized by a cutoff and the
 limit $R\longrightarrow\infty$ is subtracted from the result and finally,
 the regularization removed. The result is finite, giving 
\begin{equation}
\langle 0_R|T_{00}|0_R\rangle_{ren}=-{\pi\over{6R^2}} \label{eq2}
\end{equation}\\
\indent One convenient regularization is the zeta function regularization
(see \cite{birrel}),
 in which the expression $\sum_{m=0}^{\infty}m$ is identified with the 
 analytically continued value of the Reimann zeta function $\zeta(s)$
 at the point $s=-1$. This is finite, and is given by $\zeta(-1)=-1/12$,
 giving the vacuum energy in eqn.(\ref{eq2})  directly, without any explicit 
 subtraction. This technique is very common in the study of quantum field 
theory in curved spacetime. This is the technique we shall employ in this
 paper.
\end{section}
\begin{section}*{Calculation Of Vacuum Energy In The ADD Model}
The calculation of the vacuum energy due to compactification will now be 
 carried out in the ADD scenario. We shall consider a spacetime which is
a direct product of a non-compact 4-dimensional spacetime and two compact
 spatial extra dimensions compactified to a 2-torus. The matter fields forever
 reside on the 3-brane (the non-compact 4-dimensional spacetime) and gravity
 propagates in the bulk. The vacuum metric is Minkowskian. We consider the
 propagation of linearized gravity in this bulk spacetime. The Lagrangian
 for the system is the Fierz-Pauli Lagrangian\cite{tao} given by
\begin{equation}
{\mathcal{L}}={1\over{4}}(\partial^{\hat\mu}{\hat h}^{{\hat\nu
}{\hat{\rho}}}
\partial_{\hat\mu}{\hat h}_{{\hat\nu}{\hat\rho}}-
\partial^{\hat{\mu}}\hat{h}\partial_{\hat{
\mu}}\hat{h}-2{\hat{h}}^{\hat{\mu}}{\hat{h}}_{hat\mu}+2{\hat{h}}^{\hat{\mu}}
\partial_{\hat{\mu}}\hat{h})
\end{equation}
where ${\hat h}={\hat h}^{\hat\mu}_{\hat\mu}\,,{\hat h}_{\hat\nu}=\partial
^{\hat\mu}{\hat h}_{{\hat\mu}{\hat\nu}}$ and the Greek indices run over the
 6 bulk coordinates. Raising and lowering of indices has been carried out
 using the bulk Minkowski metric $\eta^{{\hat\mu}{\hat\nu}}=diag
(1,-1,-1,-1,-1,-1)$. The equations of motion in the de Donder gauge are
 the d'Alembert equations
\begin{equation}
\Box_{(4+2)}\Bigg({\hat h}_{{\hat\mu}{\hat\nu}}-{1\over{2}}\eta_{{\hat\mu}{
\hat\nu}}{\hat h}\Bigg)=0
\end{equation}\\
\indent Now, a Kaluza-Klein reduction is carried out as (see \cite{tao})
\begin{equation}
{\hat h}_{{\hat\mu}{\hat\nu}}={1\over{R}}\Bigg (\begin{array}
{cl}
h_{\mu\nu}+\eta_{\mu\nu}\phi & A_{\mu i}\\
A_{\nu j} & 2\phi_{ij}
\end{array}\Bigg )\label{eq3}
\end{equation}
where $R$ is the size of each extra spatial dimension, $\phi=\phi_{ii}$,
 $\mu\,,\nu=0,1,2,3$ and $\i\,,j=1,2$. The bulk linearized field can be
 expanded in terms of creation and annihilation operators. For notational
 simplicity, we shall suppress the tensor indices and the polarization 
states of the bulk graviton field and denote it by $\Phi$. Then, in terms
 of the modes, it is given by
\begin{equation}
\Phi(x,y)={1\over{R}}\sum_{n_1,n_2}\int {d^3k\over{(2\pi)
^3 2\omega(k)_{n_1,n_2}}
}[a(k)_{n_1,n_2}e^{-\imath k.x}e^{\imath{2\pi\over{R}}(n_1y_1+n_2y_2)}+h.c.]
\end{equation}
where $x$ denotes the four brane spacetime coordinates and $y$ are the two
 compact coordinates. The commutation relations satisfied by the operators
 $a(k)_{n_1,n_2}$ are 
\begin{equation}
[a(k)_{n_1,n_2},a(k')_{m_1,m_2}]=\delta_{n_1m_1}\delta_{n_2m_2}2(2\pi)^3
\omega(k)_{n_1,n_2}\delta(\vec{k}-\vec{k'})
\end{equation}
where $\omega(k)_{n_1,n_2}=\sqrt{{\vec{k}}^2+4\pi^2(n^2_1+n^2_2)/R^2}$.\\
\indent The vacuum state is defined as $a(k)_{n_1,n_2}|0_R\rangle=0$.
The vacuum expectation value of the stress-tensor associated 
 with the field $\Phi$ is calculated and identifying the quantity $R\Phi$
 as the effective four dimensional field (analogous to eqn.(\ref{eq3}))
, the four dimensional vacuum
 energy density is given by
\begin{equation}
\langle 0_R |T_{00}|0_R\rangle=
\langle E_{vac}\rangle=\sum_{n1,n2=1}^{\infty}\int {d^3k\over{(2\pi)^3}}
\sqrt{{{\vec{k}}^2+{4\pi^2(n^2_1+n^2_2)\over{R^2}}}}\label{eq4}
\end{equation}\\
\indent To regularize the $k$ integration, we write the integral as
\begin{equation}
I(m_{\bf n})=\mu^{2\epsilon}\int {d^3k\over{(2\pi)^3}}
({{\vec{k}}^2+m^{2}_{\bf n}})^{{1\over{2}}-\epsilon}
\end{equation}
where $m^{2}_{\bf n}={4\pi^2(n^2_1+n^2_2)\over{R^2}}$ and $\mu$ is a
parameter with dimensions of mass. The parameter $\epsilon$ is introduced
for regularization, and will finally be taken to be zero. In this form,
 eqn.(\ref{eq4}) is seen to represent simply the zero point energy of the massive
 Kaluza-Klein tower on the brane.
 The $k$ integral is given in terms of the beta function as 
\begin{eqnarray}
I(m_{\bf n})&=&{{m^{4}_{\bf n}}\over{4\pi^2}}\Bigg
({\mu^2\over{{m^{2}_{\bf n}}}}\Bigg)^
{\epsilon}B(3/2,-2+\epsilon)\nonumber \\
&=& -{m^{4}_{\bf n}\over{32\pi^2}}\Bigg ({1\over{\epsilon}}
+2\log 2-{1\over{2}}-\log({m^{2}_{\bf n}\over{\mu^2}})\Bigg )
\end{eqnarray}
 where the final expression holds for infinitesimal $\epsilon$. Then, the 
vacuum energy is given by 
\begin{equation}
\langle E_{vac}\rangle=-{1\over{32\pi^2}}\sum_{n1,n2=1}^{\infty}
m^{4}_{\bf n}\Bigg ({1\over{\epsilon}}
+2\log 2-{1\over{2}}-\log({m^{2}_{\bf n}\over{\mu^2}})\Bigg )
\end{equation}\\
\indent The $k$ integral has been regularized, but the summation over 
modes is still to be regularized and the renormalized value obtained. 
 This operation is carried out
 using the zeta-function regularization. We identify the summations over
 $n_1$ and $n_2$ as analytically continued values of the zeta-function
 and it's derivatives. Then,
\begin{eqnarray}
\sum_{n1,n2=1}^{\infty}m^{4}_{\bf n}&=&\Bigg({4\pi^2\over{R^2}}\Bigg)^2   
\sum_{n1,n2=1}^{\infty}(n^{4}_{1}+n^{4}_{2}+2n^{2}_{1}n^{2}_{2})\nonumber \\
&=&\Bigg({4\pi^2\over{R^2}}\Bigg)^{2}
  2[\zeta(0)\zeta(-4)+\zeta^{2}(-2)] \nonumber \\
&=& 0
\end{eqnarray}
since $\zeta(-2m)=0$, if $m$ is a natural number. Therefore, the resulting
 vacuum energy is given by
\begin{eqnarray}
 \langle E_{vac}\rangle&=&{1\over{32\pi^2}}\sum_{n1,n2=1}^{\infty}
m^{4}_{\bf n}\log({m^{2}_{\bf n}\over{\mu^2}})\\ \nonumber
&=& {\pi^2\over{2R^4}}\sum_{n1,n2=1}^{\infty}(n^{2}_{1}+n^{2}_{2})^2
\log(n^{2}_{1}+n^{2}_{2})\nonumber \\
&=& -{\pi^2\over{R^4}}\zeta(0)\,\zeta'(-4)
\end{eqnarray}
 where $\zeta(0)=-1/2$ and $\zeta'(-4)=0.008$.
\indent Since we had suppressed the graviton polarization indices in
making this calculation, we multiply the final result with the number
 of degrees of freedom for a graviton in $4+2$ dimensions, which is $9$.
Therefore, the final vacuum energy is given by
\begin{equation}
 \langle E_{vac}\rangle_{4+2}=-{9\pi^2\over{R^4}}\zeta(0)\,\zeta'(-4)
\end{equation}\\
\indent A straightforward calculation genaralizes the  result 
  to $4+n$ dimensions, and is given by
\begin{equation}
 \langle E_{vac}\rangle_{4+n}
=-{\pi^2\over{R^4}}\Bigg[{(2+n)(3+n)\over{2}}-1\Bigg]
[\zeta(0)]^{n-1}\,\zeta'(-4)
\end{equation}
where the factor in the brackets is just the number of graviton degrees
 of freedom in $4+n$ dimensions.
\end{section}
\begin{section}*{Constraints On The ADD Model}
 If the vacuum energy
 associated with the massive Kaluza-Klein modes is to account for the
 origin of the cosmological constant, it would put severe constraints
 on the number of extra dimensions and the compactification radius 
 of the extra dimensions. Firstly, it is to be noted that the sign of the
 vacuum energy depends on the number of the extra dimensions, being
 positive for even and negative for odd number of 
 extra dimensions. We would like
 to put bounds on $n$ and $R$ in the above expression by demanding that
\begin{equation}
\langle E_{vac}\rangle_{4+n}\leq 10^{-47} (GeV)^4
\end{equation}\\
\indent Then, it is seen that the bound on $R$ is not very sensitive
 to the number of extra dimensions, $n$. It is seen that for $n$ varying
 from $2$ to $10$, the above inequality approximately constrains $R$ to
 essentially the same lower value, $R\geq 0.1$ mm. However, there is a 
consistency condition relating the 4-dimensional Planck scale $M_{pl}$
 and the bulk scale of gravitation, $M_s$. This relation goes as \cite{tao}
\begin{equation}
 M^{2}_{pl}\sim R^{n}M^{n+2}_s
\end{equation}
 For $R$ constrained to $R\geq 0.1$ mm., this consistency equation gives
 sensible values of $M_s$ for $n=1,2$ only. For $n=2$, this gives $M_s\leq 1$
 TeV, which is acceptable and interesting from the hierarchy problem point
 of view. For $n=3$, $M_s$ is constrained to lie below a few GeV, which
 is phenomenologically ruled out. Higher values of $n$ become more and more
 unacceptable. The situation for $n=1$ gives $M_s\leq 10^8$ Gev, the equality
 holding for $R=0.1$ mm, and the inequality for $R> 0.1$mm. Therefore, since
 $R>0.1$ mm. will be in conflict with macroscopic gravitational experiments, 
  this situation is acceptable only for $M_s\sim 10^8$ GeV, which would however
 not get rid of the hierarchy problem.
  
$n=2$ therefore seems to be the only viable
  possibility which would solve the hierarchy problem
 and not be in conflict with high-energy experiments at the same time. However,
 since the lower constraint on $R$ is of the order of $0.1$ mm., this simply
 
 means that macroscopic tests of gravitation ought to reveal deviations
 from Newton's law of gravitation at distances comparable to this. Thus,
 if the ADD model is the correct description of spacetime and if this vacuum
 energy is to contribute to the observed cosmological constant, then 
  deviations from Newtonian gravity at scales of the order of 
 $0.1$ mm. should show up.
 Any failure in observing such deviations at length scales $\geq 0.1$ mm.
 will imply that unless there exists some deeper symmetry/mechanism that
 removes this finite vacuum energy associated with compactification, the
 ADD model fails to give an accurate description of spacetime.
\end{section}
\begin{section}*{Conclusions}
 We have calculated the topological vacuum energy of linearized
 gravity in the ADD scenario and shown that if this is to account for
 the observational bounds on the cosmological constant, the size of 
the compact dimensions should be $\geq 0.1$ mm. The 
vacuum energy is shown to be just the zero point energy of the massive 
Kaluza-Klein modes.
 The calculation has been carried out for a toroidal topology of the
 compact manifold, but the essential features of the calculation are 
not expected to change if some other locally flat non-trivial topology
 is taken. The result after zeta-function regularization and appropriate
 renormalization condition is seen to be finite, and not sensitive to
the UV cut-off of the theory, as would have been dimensionally expected.
 The reason is the subtraction of infinite (or large but finite in case
 of a finite UV cut-off) vacuum energy associated with the trivial
 topology when the extra dimensions are not curled up or compact.
 The result has been generelized for the case of an n-torus. If this
 vacuum energy is to account for observational constraints on the cosmological
 constant, this would put severe constraints on the volume of the compact 
 manifold and it's dimensionality.
\end{section}

\end{document}